%% file: draft.tex
\documentstyle[referee]{mn}
\newif\ifAMStwofonts
\def\lesssim{\mathrel{\mathpalette\oversim<}}
\def\gtrsim{\mathrel{\mathpalette\oversim>}}
\def\oversim#1#2{\lower0.2ex\vbox{\baselineskip0pt\lineskip0pt
  \lineskiplimit0pt\ialign{$#1\hfil##\hfil$\crcr#2\crcr\sim\crcr}}}
\vfill
\vspace{3cm}
\input mnmacro.tex

\title[]{The effect of bias and redshift distortions on a geometric test
for the cosmological constant}
\author[]{K.~Yamamoto$^1$, H.~Nishioka$^1$, and A.~Taruya$^2$\\
   $^1$Department of Physics, Hiroshima University,
   Kagamiyama, Higashi-Hiroshima, 739-8526, Japan\\
   $^2$Department of Physics, University of Tokyo, 
            Tokyo, 113-0033, Japan
}
\date{
in original form 2000 June}
\pagerange{\pageref{firstpage}--\pageref{lastpage}}
\pubyear{2000}

\def\LaTeX{L\kern-.36em\raise.3ex\hbox{a}\kern-.15em
    T\kern-.1667em\lower.7ex\hbox{E}\kern-.125emX}

\begin{document}
\maketitle
\label{firstpage}

\begin{abstract}
We revisit the feasibility of a cosmological test with the geometric 
distortion focusing on an ambiguous factor of the evolution of bias.
Starting from defining estimators for the spatial two-point correlation 
function and the power spectrum, in a rigorous manner, we derive useful 
formulas for the two-point clustering statistics which take the 
light-cone effect and the redshift-space distortions into account. 
Then we investigate how the predicted correlation functions are 
affected by the redshift-space distortions and the bias assuming 
quasar samples which roughly match the 2dF survey. Effect of the 
bias is crucial, in particular, on anisotropic component of 
the clustering statistics in redshift-space. By surveying 
behavior of the predicted correlation functions to parameters of 
a phenomenological model for the evolution of bias, it is shown 
that the correlation functions on the light-cone is sensitive to 
a mean amplitude of the bias and rather insensitive to the speed of 
its redshift-evolution. This feature is useful for an analysis
of the geometric test.

\end{abstract}


\vspace{1cm}
\newpage

\section{Introduction}

The wide-field redshift survey projects, Two-degree Field (2dF) and 
Sloan Digital Sky Survey (SDSS), are upcoming, and large numbers of 
quasars and galaxies are expected to be detected.
In particular, the 2dF group has recently reported their preliminary 
results about the evolution of the quasar luminosity function and the
spatial correlation function based on the several thousands quasars 
detected from the survey so far (Shanks et~al. 2000; Boyle et~al. 2000).
\footnote{After we have completed the present manuscript, the most recent
result on the 2dF quasar survey has been reported (Croom et al. 2000), 
in which a clustering analysis of the quasars is presented in detail. }
The SDSS survey project will also provide a set of high precision data 
for the distribution of quasars in near future.
These surveys will provide very important clues on the origin and the 
evolution of the quasars, and will put constraints on theoretical 
models of the structure formation of the universe. However, there 
seems to remain room for discussion as to how we compare the
observational data with theoretical models and what information is drawn 
out from them. Concerning the latter subject, various cosmological tests 
with the future data have been proposed. Among such cosmological tests, 
the geometric test using the cosmological 
redshift-space (geometric) distortion is quite unique. 

Alcock and Paczynski first pointed out the possibility of the geometric 
test (1979). They pointed out that if spherically
distributed structures of cosmological objects were distributed 
in the universe, the observation of such structures at various 
redshift in redshift space offers a plausible test for the 
cosmological model of the universe, which is sensitive to a 
cosmological constant. 
This test is based on the fact that the shape of the spherically 
distributed structure is observed distorted in the cosmological 
redshift-space, where an incorrect distance-redshift relation 
is assumed. We refer to the distortion as the cosmological redshift-space 
(geometric) distortion. The geometric distortion
is traced back to a coordinate transformation of the distance 
between real space and redshift space.
Several authors have extensively discussed about possible
tests for cosmological models with the geometric distortion
with the clustering statistics of high-redshift objects 
(Ryden 1995;
Matsubara \& Suto 1996; Ballinger et~al. 1996; Popowski et~al. 
1998; Nair 1999; Hui et~al. 1999). 

In principle, the geometric test is a unique test for the 
cosmological model, however, a measurement of the 
geometric distortion suffers from various other 
observational effects. In particular the peculiar motion 
of the cosmological objects causes additional distortion 
in the distribution in redshift space. 
Especially the linear distortion, which is due to
the bulk motion of the cosmological objects, can be influential 
even at high-redshift on large scales. The net of the
effect depends on the amplitude of the bias. 
Furthermore, the light-cone effect would become an important effect 
when the statistical quantities like the correlation function and 
the power spectrum are computed using the data in a wide range of 
the redshift. In this case, the evolution of the bias is an
important but ambiguous factor in predicting the correlation 
function and the power spectrum (Suto, Magira, Yamamoto 2000; 
see also references therein). 
The theoretical studies based on the Press-Schechter theory and 
the extensive numerical works recently show that the time evolution of 
clustering bias for the galaxies and the clusters of galaxies 
strongly depends on the cosmological models (e.g, Mo \& White 1996; 
Blanton et al. 1999; Somerville et~al. 2000; Taruya \& Suto 2000),  
however, the quantitative 
prediction for the evolution of galaxy clustering is still 
under consideration. As for the quasars, the uncertainty of the 
formation mechanism makes for further difficulty in predicting 
the evolution of bias. 

In the present paper we consider the details of the theoretical predictions 
for quasar two-point statistics focusing on the evolution of the bias 
factor. We discuss how we can draw out information
about the evolution of the bias from the two-point statistics of 
cosmological objects and how the evolution of bias 
alters the prediction for the two-point statistics, which determines
feasibility of the cosmological test with the geometric distortion.
This paper is organized as follows: In section 2,
we summarize theoretical formulas for the two-point correlation
function and the power spectrum which incorporate the light-cone effect
and the redshift-space distortion effects. The final expression 
of the formula has been presented in reference (Suto, Magira, Yamamoto 
2000). On the contrary to this previous paper, we here derive the
theoretical formulas in a rigorous manner by defining estimators for
the two-point correlation function and the power spectrum on a light 
cone. This makes clear how our theoretical two-point correlation 
function and power spectrum are related to the estimators in 
data processing. Then we investigate the details of the correlation 
functions assuming quasar samples which roughly match the 2dF survey 
in section 3. Section 4 is devoted to summary and discussions. Throughout 
this paper we use the unit in which the light velocity $c$ equals $1$.

\section{Formalism}
\newcommand{\cpara}{c_{\scriptscriptstyle \|}}
\newcommand{\cperp}{c_{\scriptscriptstyle \bot}}
\def\bfR{{\bf R}}
\def\bfk{{\bf k}}
\def\vgamma{{\vec\gamma}}
\def\s{{s}}
\def\S{{S}}
\def\bfs{{\bf s}}
\def\bfS{{\bf S}}
\def\bfR{{\bf R}}
\def\P{L}

In this section we summarize formulas for the two-point statistics 
of the distribution of cosmological objects in redshift space 
incorporating the light-cone effect. On the contrary to the previous 
paper by Suto, Magira, \& Yamamoto (2000), in which the final expression
for the two-point correlation function is presented, we here present 
a review to derive the formula starting from 
estimators for the anisotropic correlation function and the 
power spectrum in cosmological redshift-space. 

We start from reviewing an idealized data processing to compute the 
two-point correlation function and the power spectrum from a map of
cosmological objects. Assuming a redshift survey, data is given 
in terms of a set of pairs of the direction vector $\gamma$ and the 
redshift $z$ of the cosmological objects.
When constructing the three dimensional map of the cosmological 
objects, we must introduce a radial coordinate $s$ by assuming 
some distance-redshift relation $s=s(z)$. Because the 
cosmological model of our universe is not definitely determined, 
we must {\it assume} the distance-redshift relation $s=s(z)$.
Frequently the distance extrapolating the Hubble law, $s=s_{\rm Hb}(z)$, 
or the comoving distance in the Einstein de~Sitter universe, 
$s=s_{\rm Ed}(z)$, have been used for the relation, where we defined
\begin{eqnarray}
&&s_{\rm Hb}(z)={z\over H_0},
\\
&&s_{\rm Ed}(z)={2\over H_0}(1-1/\sqrt{1+z}),
\end{eqnarray}
with the Hubble constant $H_0=100 h {\rm km/s/Mpc}$. 
We refer to the three dimensional space $(s,\gamma)$ as the 
cosmological redshift-space.
For a sample at high-redshift $z\gtrsim1$, extrapolation of
the Hubble law, $s=s_{\rm Hb}(z)$, is unphysical. However,
for comparison, we consider the both $s=s_{\rm Hb}(z)$ and 
$s=s_{\rm Ed}(z)$ in sections 3.1 and 3.2, according 
to the previous investigations (Matsubara \& Suto 1996; 
Ballinger et~al. 1996). After section 3.3 we adopt 
$s=s_{\rm Ed}(z)$.

Next we assume that the number density field $n(\bfs)$ with $\bfs=s\gamma$ 
can be constructed from the map. By introducing a synthetic catalogue 
$n_{\rm syn}(\bfs)$, we construct the density field:
\begin{eqnarray}
  F(\bfs)={\displaystyle{n(\bfs)-n_{\rm syn}(\bfs)}
  \over \displaystyle{[\int d\bfs \tilde n(\s)^2]^{1/2}}},
\end{eqnarray}
where $\tilde n(s)$ denotes the mean number density at
the distance $s$. With the use of $\tilde n(s)$, we write
\begin{eqnarray}
 &&n(\bfs)=\tilde n(s)\bigl(1+\Delta(\bfs)\bigr),
\\
 &&n_{\rm syn}(\bfs)=\tilde n(\s)\bigl(1+\Delta_{\rm syn}(\bfs)\bigr).
\end{eqnarray}
The synthetic field $n_{\rm syn}(\bfs)$ satisfies 
$\bigl<n_{\rm syn}(\bfs_1)n_{\rm syn}(\bfs_2)\bigr>=
\bigl<n(\bfs_1)n_{\rm syn}(\bfs_2)\bigr>=\tilde n(\s_1) \tilde n(\s_2)$,
which lead
\begin{eqnarray}
  \bigl<\Delta_{\rm syn}(\bfs_1)\Delta_{\rm syn}(\bfs_2)\bigr>=
  \bigl<\Delta_{\rm syn}(\bfs_1)\Delta(\bfs_2)\bigr>=0.
\label{synth}
\end{eqnarray}

Then the two-point correlation function can be computed by
averaging the products of $F(\bfs)$ over all the possible 
pairs of two-points separated by the distance $R$. 
A mathematical expression for this process
can be explicitly written 
\begin{eqnarray}
 &&\xi_l(R)=\int  
 {d\Omega_{\hat\bfR}\over 4\pi}\int d\bfs_1 \int d\bfs_2
 F(\bfs_1)F(\bfs_2)
   \delta^{(3)}(\bfs_1-\bfs_2-\bfR)\P_l(\hat\bfs_h\cdot\hat\bfR)(2l+1),
\label{defxil}
\end{eqnarray}
where $R=|\bfR|$, $\hat\bfR=\bfR/R$, $\bfs_h=(\bfs_1+\bfs_2)/2$, 
$\s_h=|\bfs_h|$, $\hat\bfs_h=\bfs_h/\s_h$, and $\P_l(\mu)$ is 
the Legendre polynomial of the $l$-th order. (see figure 1.)
In the case $l=0$, where $\P_0(\mu)=1$, $\xi_0(R)$ represents
the isotropic (angular averaged) correlation function.
On the other hand, other cases measure the anisotropies
of the correlation function owing to the factor
$\P_l(\hat\bfs_h\cdot\hat\bfR)$. 
In the particular case $l=2$, $\xi_2(R)$ expresses the quadru-pole anisotropy 
of the correlation function. We adopt this estimator
for the isotropic and the anisotropic correlation functions. 

Similar to the case of the correlation function we define
the following estimator for the power spectrum (cf., Yamamoto, Nishioka, 
\& Suto 1999)
\begin{eqnarray}
  &&P_l(k)=\int 
 {d\Omega_{\hat\bfk}\over 4\pi}\int d\bfs_1 \int d\bfs_2
 F(\bfs_1)F(\bfs_2)
   e^{i\bfk\cdot(\bfs_1-\bfs_2)}\P_l(\hat\bfs_h\cdot\hat\bfk)(2l+1),
\end{eqnarray}
where $k=|\bfk|$, $\hat\bfk=\bfk/k$, and $\bfk$ is the wave
number vector.
When adopting the above estimators for the correlation function 
and the power spectrum, which we have defined independently,
we can show that a familiar relation holds between them.
Namely, by using the mathematical formulas,
\begin{equation}
  \delta^{(3)}(\bfs_1-\bfs_2-\bfR)={1\over (2\pi)^3}\int d\bfk
  e^{i\bfk\cdot(\bfs_1-\bfs_2-\bfR)},
\end{equation}
and
\begin{equation}
  e^{i\bfk\cdot\bfR}=4\pi \sum_{l,m}i^lj_l(kR)
  Y_l^m{}^*(\Omega_{\hat\bfk})Y_l^m(\Omega_{\hat\bfR}),
\end{equation}
where $Y_l^m(\Omega)$ and $j_l(x)$ denote the spherical harmonics 
on an unit sphere and the spherical Bessel function, respectively,
we can verify that the following relation holds
\begin{equation}
  \xi_l(R)={1\over2\pi^2 i^l}\int dk k^2 j_l(kR) 
  P_l(k).
\label{Fourie}
\end{equation}
The inverse transformation leads
\begin{equation}
  P_l(k)={4\pi i^l}\int dR R^2 j_l(kR) 
  \xi_l(R).
\label{invtr}
\end{equation}

As is well recognized as the problem of the cosmic variance,
expectation values of the estimators are only predictable
theoretically. Then we consider the ensemble average of 
the estimators, i.e.,
\begin{eqnarray}
  \xi_l^{\rm LC}(R)&=&\bigl<\xi_l(R)\bigr>,
\\
  P_l^{\rm LC}(k)&=&\bigl<P_l(k)\bigr>,
\end{eqnarray}
which define the correlation function and the power
spectrum on a light cone.

In Appendix we show that, applying the small angle approximation, 
$P_l^{\rm LC}(k)$ is reduced to
\begin{equation}
  P^{\rm LC}_l(k)={\displaystyle{\int dz \Bigl({dN\over dz}\Bigr)^2
                      \Bigl({s^2 ds\over dz}\Bigr)^{-1} P_l^{\rm crd}(k,z)}
                \over \displaystyle{\int dz \Bigl({dN\over dz}\Bigr)^2
                      \Bigl({s^2 ds\over dz}\Bigr)^{-1}}},
\label{Pcllk}
\end{equation}
where $dN/dz$ denotes the number count of the objects per unit redshift
and per unit solid angle, $P_l^{\rm crd}(k,z)$ is defined:
\begin{eqnarray}
  &&P_l^{\rm crd}(k,z)={2l+1\over \cperp^2\cpara}\int_0^1 d\mu 
  \P_l(\mu)b(q[k],z)^2P_{\rm mass}(q[k],z)
\nonumber
\\
  &&{\hspace{0.0cm}}\times
  \biggl\{
  {1+\mu^2((1+\beta(z))\omega^2-1)\over 1+\mu^2(\omega^2-1)}\biggr\}^2
  D\biggl[{\mu k\sigma_p(z)\over \cpara}\biggr],
\label{Plcrd}
\end{eqnarray}
where $P_{\rm mass}(q,z)$ denotes the power spectrum of the total mass distribution 
in real space with the wave number $q$,     
$D[{\mu k\sigma_p(z)/\cpara}]$ describes the damping function
due to the Finger of God effect, and with denoting the comoving 
radial distance in real space $r(z)$ and the distance in 
cosmological redshift-space $s(z)$ (see equations (1), (2) and (A27) 
for the definitions), we defined 
\begin{eqnarray}
  &&\cperp={r(z)\over s(z)},
\label{defcperp}
\\
  &&\cpara={dr(z)\over ds(z)},
\\
  &&\omega={\cperp\over \cpara},
\\
  &&q[k]=k\sqrt{{(1-\mu^2)\over\cperp^2}+{\mu^2\over \cpara^2}},
\end{eqnarray}
and 
\begin{eqnarray}
  &&\beta(z)={1\over b(q[k],z)}{d \ln D_1(z) \over d \ln a(z)},
\label{betafactor}
\end{eqnarray}
with the linear growth rate $D_1(z)$ and the scale factor $a(z)$. 
The function $b(q,z)$ denotes the linear bias factor defined in 
real space (see Appendix).
Expression (\ref{defcperp}) is correct only for the case when the real space 
is spatially flat. (see equation (\ref{defcperpp}) for 
generalization to an open universe)
The correlation function $\xi_l^{\rm LC}(R)$ 
can be computed from $P_l^{\rm LC}(k)$ with the use of equation 
(\ref{Fourie}).

The expression $P_l^{\rm crd}(k,z)$ is well-known as a power spectrum 
on a constant time hypersurface in the cosmological redshift-space 
(e.g., Ballinger et~al. 1996; Suto et~al. 1999; Magira, Jing, \& Suto 
2000). The term including the $\beta(z)$-factor in the expression 
$P_l^{\rm crd}(k,z)$ is originated from the Kaiser factor which 
represents the effect of the linear distortion (Kaiser 1987). 
In the present paper we adopt the exponential model for the distribution 
of the pairwise peculiar velocity, then the damping factor is 
\begin{eqnarray}
  D[{\mu k\sigma_p(z)/\cpara}]={1\over 1+\mu^2k^2\sigma_p(z)^2/2\cpara^2},
\end{eqnarray}
where $\sigma_p(z)$ is the pairwise velocity dispersion. For 
$P_{\rm mass}(q,z)$, the power spectrum in real space at the redshift $z$, 
we adopt the linear power spectrum with the CDM transfer function 
(Bardeen et~al. 1986) and the Peacock \& Dodds formula for the nonlinear 
power spectrum (Peacock \& Dodds 1994; 1996). The essence of the 
geometric distortion is only the coordinate transformation 
between $(s(z),\gamma)$ and $(r(z),\gamma)$, which is described by 
the coefficients $\cperp$ and $\cpara$.  

Equation (\ref{Pcllk}) indicates that the light-cone effect is 
incorporated by averaging the local power spectrum $P_l^{\rm crd}(k,z)$
by weighting the factor in the formula:
\begin{equation}
  W(z)=\biggl({dN\over dz}\biggr)^2 \biggl(s^2{ds\over dz}\biggr)^{-1}.
\end{equation}
The formula (\ref{Pcllk}) is obtained under the small angle approximation.
Hence the validity of the use is limited to small scales. For large scales,
$R\gtrsim s_{\rm max}$, where $s_{\rm max}$ is the size of a survey area, 
the finite size effect of the survey area affects the proper estimation of
the correlation function and the power spectrum. 
(e.g., Yamamoto \& Suto 1999; Nishioka \& Yamamoto 1999; 2000). 

\section{QSO Two-point Statistics and the Evolution of Bias}
\def\rmB{{\rm B}}
In this section we discuss the detail of the two-point correlation 
function focusing on how the quasar two-point statistics depend on
the evolution of bias. Motivated from the recent report by 
Shanks et~al. (2000), we assume a sample which roughly corresponds 
with the 2dF quasar survey.

\subsection{The Quasar Luminosity Function}

For simplicity we here adopt the B-band quasar luminosity function 
according to Wallington \& Narayan (1993). We set the limiting magnitude 
$B_{\rm lim}=20.85$, which is useful to obtain a sample which roughly 
matches the 2dF quasar survey. According to Wallington \& Narayan 
(1993; see also Nakamura \& Suto 1997), we adopt
\begin{eqnarray}
  &&\Phi(M_\rmB,z)=\Phi^*\times\bigl[
  10^{0.4[M_\rmB-M_\rmB(z)](\kappa_1+1)}
  +10^{0.4[M_\rmB-M_\rmB(z)](\kappa_2+1)}\bigr]^{-1}, 
\end{eqnarray}
where
\begin{eqnarray}
  &&\Phi^*=6.4\times10^{-6}~h^{3}{\rm Mpc^{-3}}{\rm mag}^{-1},
\\
  &&M_\rmB(z)=M_\rmB^*-2.5k_L\log(1+z),
\\
  &&M_\rmB^*=-20.91+5\log h,
\end{eqnarray}
with $k_L=3.15$, $\kappa_1=-3.79$, $\kappa_2=-1.44$, which is
applicable for $0.3<z<2.2$.
The B-band apparent magnitude is computed from a quasar of 
absolute magnitude $M_\rmB$ at $z$,
\begin{equation}
  B=M_{\rmB}+5\log\biggl[{d_L(z)\over 10 
  {\rm pc}}\biggr]-2.5(1-p)\log(1+z),
\end{equation}
where $d_L(z)$ is the luminosity distance and we adopt the quasar 
energy spectrum $L(\nu)\propto \nu^{-p}$ (p=0.5) for the K-correction. 
Because the luminosity function is constructed under the assumption of 
the Einstein de Sitter universe, we here use the luminosity distance 
\begin{equation}
  d_L(z)={2\over H_0}(1+z-\sqrt{1+z}),
\end{equation}
which holds in the Einstein de Sitter universe.

The number count of the sample with the B-band limiting magnitude 
$B_{\rm lim}$ per unit redshift and unit solid angle is expressed
\begin{equation}
  {dN\over dz}= {r^2 dr \over dz} 
  \int_{M_{\rmB_{\rm lim}}}^\infty dM_\rmB\Phi(M_\rmB,z),
\end{equation}
with $r$ being the comoving distance. Because of the same reason for the 
luminosity distance $d_L$, we adopt the comoving distance in the 
Einstein de Sitter universe being $r=d_L(z)/(1+z)$.
Figure 2 plots the weight factor $W(z)$ as the function 
of redshift with adopting cosmological redshift-space
extrapolating the Hubble law $s=s_{Hb}(z)$ (dotted line)
and Einstein de Sitter universe $s=s_{Ed}(z)$ (solid line). 
From this figure we see that the choice of the cosmological
redshift-space significantly alters the weight factor.
After completing almost of the present paper, the preliminary report 
on the luminosity function from the 2dF quasar survey 
(Boyle et al. 2000) has been announced. 
They have found the consistency of the luminosity function 
with the previous results. The use of the new luminosity 
function does not alter our conclusions described below.

\subsection{QSO Spatial Clustering and Evolution of Bias}
Recently bias models for the quasar distribution have been proposed 
(e.g., Martini \& Weinberg 2000; Haiman \& Hui 2000; Fang \& Jing 1998).
Construction of theoretical models for the quasar bias is a challenging 
problem, and we need more investigations. However, the bias mechanism of 
the quasar distribution contains many uncertain factors at present.
In the present paper we consider one of the phenomenologically simplest 
models of the bias with the form, 
\begin{equation}
  b(z)=\alpha+\bigl(b(z_*)-\alpha\bigr)\biggl({1+z\over 1+z_*}\biggr)^\beta,
\label{defbz}
\end{equation}
where $\alpha$, $\beta$, and $b(z_*)$ are the constant free-parameters
(cf., Matarrese et~al. 1997). 
Throughout the present paper we use $\beta$ to denote the
free parameter of the bias evolution, which should be distinguished 
from $\beta(z)$ which we use to denote $\beta(z)$-factor for the 
linear distortion. In equation (\ref{defbz}), $\beta$ specifies the rate of
evolution of bias, $b(z_*)$ does the amplitude of bias at 
some redshift $z_*$, and $\alpha$ corresponds to the amplitude
of bias at $z=0$ in the limit $z_*\gg1$. 
We have assumed the bias model is independent of scales $k$.
This assumption relies on the recent numerical simulations 
that the scale-independent bias can be valid on the 
linear and quasi-linear scales as long as the mechanism of bias 
is dominantly affected by the gravity (e.g, Mann, Peacock \& Heavens
1998). 
Although the realistic bias model taking account of the nonlinearity 
and stochasticity should be considered correctly (Dekel \& Lahav 1999;
Fry 1996; Tegmark \&
Peebles 1998; Taruya, Koyama \& Soda 1999; Taruya \&
Suto 2000), the deterministic linear bias parameterized by 
the one parameter $b$ can become a good approximation even in the 
quasi-linear regime for galaxies and clusters (Taruya et al. 2001; 
see also Scherrer \& Weinberg 1998). 

As noted above, both of the power spectrum $P^{\rm LC}_l(k)$ 
and the correlation function $\xi^{\rm LC}_l(R)$, in principle, 
carry the same information because they are connected by the Fourier
transformation. As the light-cone effect can be regarded as an
averaging process, therefore, it is important to understand the 
evolution of $P^{\rm crd}_l(k,z)$ (or $\xi^{\rm crd}_l(R,z)$) at 
different redshift.

Figure 3 shows typical behaviors of the power spectrum $P^{\rm crd}_0(k,z)$
and $P^{\rm crd}_2(k,z)$ at different redshifts adopting the 
cosmological redshift-space with $s=s_{Hb}(z)$. 
The upper panels show $P^{\rm crd}_0(k,z)$, the middle panels show
$P^{\rm crd}_2(k,z)$, and the lower panels show the ratio 
$P^{\rm crd}_2(k,z)/P^{\rm crd}_0(k,z)$.
The left panels show the case for $z=0.5$, the center for $z=2.0$, and the 
right for $z=4$. Here we adopted the bias model given by equation (\ref{defbz})
with $\alpha=0.5, \beta=1, b(z_*=2)=2$, and the real space of the CDM model 
with a cosmological constant, to be specific, 
$\Omega_0=0.3$, $\Omega_\Lambda=0.7$, $h=0.7$, $\sigma_8=1.0$.
The $\Lambda$CDM model reproduces the observed cluster abundance
(Kitayama \& Suto 1997).
Figure 4 is same as figure 3, but is the case adopting the 
cosmological redshift-space with $s=s_{Ed}(z)$.

In each panel, the dashed line plots the case when the distortion due 
to the peculiar motion of the sources are neglected, to be specific,
the Kaiser factor, the nonlinear and the finger of god effect are
neglected in the formula. Hence the dashed line shows the case when only 
the geometric  distortion is taken into account. The dotted line 
shows the case when the nonlinear and the finger of God effects are 
neglected, and the solid line plots the case obtained from the 
formula (15), in which all the effects are incorporated.  These two 
figures 3 and 4 indicate how the three kinds of the distortion effects, 
i.e., the geometric distortion, the linear distortion, the finger of 
god effect, are effective at different redshifts and on different 
scales. 
The three lines, the dashed line, the dotted line, and the solid line,
in upper panels behave in a similar way. Therefore, the effect from the 
peculiar motion is small for the amplitude $P_0^{\rm crd}(k,z)$. 
The geometric distortion significantly
affects the amplitude of the power spectrum $P_0^{\rm crd}(k,z)$.
Note also that the effect acts in different ways 
for the cases $s=s_{Hb}(z)$ and $s=s_{Ed}(z)$. As the redshift becomes
higher, the amplitude of $P_0^{\rm crd}(k,z)$ becomes larger (smaller) 
in the case with $s=s_{Hb}$ ($s=s_{Ed}$). 
This difference is caused by the scaling effect through the 
coefficients $\cperp$ and $\cpara$ (See also Taruya \& Yamamoto 2000).
Thus, for the power spectrum, $P_0^{\rm crd}(k,z)$, the geometric
distortion and the bias are the most important effects for the amplitude.

We read the following behaviors from the middle 
and lower panels of figures 3 and 4. For the anisotropic power 
spectrum, the redshift-dependence is remarkable when adopting $s=s_{Hb}(z)$.
In this case, the geometric distortion is a minor 
effect, however, the linear distortion and the finger 
of god effect are effective on the large scales and on the small 
scales, respectively,  at low redshifts. 
At the higher redshifts, the geometric distortion also becomes 
influential. On the contrary to this, in the case adopting $s=s_{Ed}(z)$,
the geometric distortion effect is a minor effect compared with the
linear distortion for $P_2^{\rm crd}(k,z)$ even at high-redshift $z=4$.
Since we here fixed the bias parameter, this feature may be strongly 
dependent on the behavior of the bias parameters. Therefore we investigate 
the bias-dependence of the correlation function in detail 
in the next subsections.

From figures 3 and 4 it is apparent that the choice of the 
cosmological redshift-space affects the behavior of the power 
spectrum (the correlation functions). 
To avoid the unphysical increase of the clustering amplitude at 
high-redshift in the case $s=s_{Hb}(z)$ and to compare with
the observational result (Shanks et~al. 2000), in which 
QSO correlation function is reported by assuming the Einstein
de Sitter universe, hereafter we adopt $s=s_{Ed}(z)$.
Though the behavior of the correlation functions depends on the choice 
of the cosmological redshift-space, sensitivity of the ratio of the 
correlation functions on a light-cone to the cosmological constant 
will not be significantly altered (see subsections 3.3 and 3.4), 
neither will be the feasibility of the geometric test.

\subsection{Characteristic Correlation Length Determines QSO Bias}
In this subsection we consider how the amplitude of the correlation 
functions which incorporate the light-cone effect depends on the evolution 
of bias. Here we determined $z_*$ of the bias model by the formula
\begin{equation}
  z_*={\int dz z W(z)\over \int dz W(z)},
\end{equation}
which describes a mean redshift. Assuming the quasar sample 
with $B_{\rm lim}=20.85$ in the range $0.3<z<2.3$, we have
$z_*=1.3$ adopting $s=s_{Ed}(z)$. We adopt this set of the values below.

Now we define the characteristic correlation length $R_c$ by
\begin{eqnarray}
  \xi_0^{\rm LC}(R_c)=1.
\label{defofRC}
\end{eqnarray}
Figure 5 shows contour of $R_c$ on the plane of the bias 
parameters $b(z_*)$ and $\beta$, which is computed from
the definition (\ref{defofRC}) with the theoretical correlation 
function $\xi_0^{\rm LC}$. Hereafter, unless otherwise stated,
the correlation functions are computed by taking into account all 
the effects, i.e., the linear distortion, the geometric distortion, 
the nonlinear and the light-cone effects described in section 2.
The left panels show the case $\alpha=0$, the center
$\alpha=0.5$, and the right $\alpha=1.0$, respectively. 
In the upper panels, the cosmological model with a cosmological constant, 
$\Omega_0=0.3, \Omega_\Lambda=0.7,h=0.7, \sigma_8=1$ is assumed as 
the model for real space, and the Einstein de Sitter universe is adopted 
as the cosmological redshift-space. The lower panels show the case 
assuming the real space of an open universe $\Omega_0=0.3, \Omega_K=0.7, 
h=0.7, \sigma_8=1$. It is notable that $R_c$ does not significantly 
depend on the parameters $\alpha$ and $\beta$, but is sensitive to 
$b(z_*)$, the amplitude of the bias at $z_*$. This feature is caused 
because $z_*$ is roughly the mean value of the redshift for averaging 
the correlation function.

Shanks et~al. (2000) have reported that the QSO correlation function 
is consistent with being $(r/r_0)^{-1.8}$ with $r_0=4h^{-1}{\rm Mpc}$. 
This results is consistent with the previous paper (Croom \& Shanks 1996), 
in which the correlation function was fitted in the similar form 
with same power index with $r_0=6h^{-1}{\rm Mpc}$. 
At present it seems that there exist not small statistical errors in
the observed correlation function of the quasar distribution. 
If the value of $R_c$ was fixed, it will put a significant constraint on
the evolution of bias. 
Though we need to assume the value of $\sigma_8$, the characteristic 
correlation length determines the QSO bias.
For example, if we seriously take the result 
$r_0=4\sim 6h^{-1}{\rm Mpc}$ by Shanks et~al. (2000) and 
Croom \& Shanks (1996), figure 5 suggests that the bias of the quasar 
sample has the amplitude of factor $1\sim 2$ at $z_*=1.3$. 
However, this statement might not be taken so seriously, because the 
result is preliminary and the sample is not homogeneous at present.
And also note that the estimators to evaluate the correlation function
should be taken in a consistent way in data processing 
according to the theoretical computations.

\subsection{Feasibility of the Geometric Test}

The geometric distortion will be detected by measuring $\xi_2(R)$, so
we next consider the anisotropic part of the correlation function, 
$\xi_2^{\rm LC}(R)$. The anisotropic power spectrum $P_2^{\rm LC}(k)$
can also measure the geometric distortion (Ballinger et al. 1996). 
In the present paper, however, we consider the 
ratio $\xi_2^{\rm LC}(R)/\xi_0^{\rm LC}(R)$ (cf., 
Matsubara \& Suto 1996). Figure 6 shows the 
contour of the ratio $\xi_2^{\rm LC}(R)/\xi_0^{\rm LC}(R)$
on the plane of the bias parameters $b(z_*)$ and $\beta$, with fixed 
$R=5h^{-1}{\rm Mpc}$ (upper panels), $R=10h^{-1}{\rm Mpc}$ (middle panels), 
and $R=20h^{-1}{\rm Mpc}$ (lower panels).
The left panels show the case $\alpha=0$, the center $\alpha=0.5$, and the 
right $\alpha=1.0$. In this figure the $\Lambda$CDM model is assumed
as the model of real space, where the same cosmological parameters are 
taken as those in figure 5, and the Einstein de Sitter universe is 
chosen for the cosmological redshift-space. 
It can be read from this figure that the dependence on $\alpha$
is weak, and that the dependence on $\beta$ becomes rather weak for 
$R\gtrsim 10h^{-1}$Mpc. The amplitude of $\xi_2^{\rm LC}/\xi_0^{\rm LC}$
is most sensitive to the parameter $b(z_*)$. The reason would be same as 
that described in subsection 3.3 for contour of $R_c$. 
 
To show the importance of the linear distortion, we show the contour 
of the ratio $\xi_2^{\rm LC}/\xi_0^{\rm LC}$ with fixing $R=10 h^{-1}$Mpc, 
in figure 7. In this figure the same model parameters as those in figure 
6 are adopted, however, the distortion effects due to the
peculiar motions are omitted on purpose in the upper and the middle panels.
The upper panels show the case only the geometric distortion is
incorporated but the distortion effect due to the peculiar motion of 
sources, i.e., the linear distortion, nonlinear and the finger of god 
effects, are neglected. 
The middle panels show the prediction within linear theory, that is, 
the case the linear distortion effect is incorporated but the 
nonlinear and the finger of God effects are omitted in the computation.
The lower panels show the case all the effects are incorporated
(Same as the middle panels in figure 6).
From figure 7 it is apparent that the middle and the lower panels are 
significantly different from the upper panels. This means 
that the linear distortion, the nonlinear and the finger of god 
effects are the dominant effects for $\xi_2^{\rm LC}$ at this scale. 
In a redshift survey, the peculiar motion of sources inevitably 
contaminate the map. These facts limit the ability of the 
cosmological test with the geometric distortion because the 
linear distortion effect is sensitive to the amplitude of bias.

Nevertheless we would like to discuss that the cosmological test
with the geometric distortion might be a useful tool for testing 
the cosmological model of our universe with the precise measurement 
of the correlation functions.
In order to discuss the feasibility of the geometric test, we also
consider the case that the real space is an open hyperbolic universe. 
Prescription to compute the correlation functions for the open 
universe is described in Appendix. Figure 8 shows the contour of 
$\xi_2^{\rm LC}(R)/\xi_0^{\rm LC}(R)$ to demonstrate the feasibility 
of the geometric test. The solid lines show contour of
$\xi_2^{\rm LC}(R)/\xi_0^{\rm LC}(R)$ with fixed $R=10h^{-1}$Mpc
for various cosmological models 
(a) $\Lambda$CDM model with $\Omega_0=0.3$;
(b) $\Lambda$CDM model with $\Omega_0=0.4$;
(c) open CDM model with $\Omega_0=0.3$;
(d) open CDM model with $\Omega_0=0.4$.
In each model we adopted $h=0.7$, $\alpha=0.5$, and $\sigma_8$ 
normalized by cluster abundance (Kitayama \& Suto 1997).
The region between the dashed lines in each panel satisfies
$4 \leq R_c\leq 6$ for each model.
Therefore $\xi_2^{\rm LC}/\xi_0^{\rm LC}$ at $R=10h^{-1}{\rm Mpc}$ 
is in the range $-0.8\sim-0.5$ for the $\Lambda$CDM model taking the 
range of the bias parameter consistent with $R_c=4\sim 6h^{-1}{\rm Mpc}$ 
into account. While  $\xi_2^{\rm LC}/\xi_0^{\rm LC} \sim -0.5\sim -0.3$ 
is predicted in the open CDM models. In this figure we fixed $\alpha=0.5$, 
however, the dependence on $\alpha$ is weak for $0\lesssim \alpha\lesssim 1$. 

Figure 9 shows the same figure 8 but with fixed $R=20h^{-1}$Mpc. 
In this case we find that $\xi_2^{\rm LC}/\xi_0^{\rm LC}$ is in 
the range $-2.0\sim-1.2$ ($-2.0\sim-1.4$) for the $\Lambda$CDM 
with $\Omega_0=0.3$ ($\Omega_0=0.4$) taking the range of the bias 
parameter consistent with $R_c=4\sim 6h^{-1}{\rm Mpc}$ 
into account. While  $\xi_2^{\rm LC}/\xi_0^{\rm LC} \sim -1.2\sim -0.8$
($-1.4\sim -1.0$) is predicted in the open CDM model with $\Omega_0=0.3$ 
($\Omega_0=0.4$). This figure demonstrates that the difference in 
$\xi_2^{\rm LC}/\xi_0^{\rm LC}$ between the $\Lambda$CDM model and 
the open model appears irrespective of the scale $R$.
At the same time this figure shows that the predictions slightly 
depend on the value of $\Omega_0$. 
If the constraint on $R_c$ becomes tight, the difference in 
$\xi_2^{\rm LC}/\xi_0^{\rm LC}$ becomes remarkable.
The ratio $\xi_2^{\rm LC}/\xi_0^{\rm LC}$ does not depend on 
$\sigma_8$ at large length scales unless the nonlinear effect 
becomes influential. However, we should remind that the assumption 
of the value of $\sigma_8$ is needed to constrain the bias 
parameters by the characteristic correlation length.

This difference in $\xi_2^{\rm LC}/\xi_0^{\rm LC}$ between 
the $\Lambda$CDM model and the open model occurs due to the
difference of the linear growth rate and the 
geometric distortion effect. The scaling effect due to
the geometric distortion alters the amplitude of $\xi^{\rm LC}_0(R)$ 
through the coefficient $\cperp$ and $\cpara$. The evolution of 
$\cperp$ and $\cpara$ in the $\Lambda$CDM model is quite different 
from those in the open model.
The main reason for the difference in $\xi_2^{\rm LC}/\xi_0^{\rm LC}$
is from the difference in $\xi_0^{\rm LC}$ due to the scaling 
effect of the geometric distortion. The difference appears for 
wide range of the parameters of the bias evolution. 
In order to check the model dependence of the bias, we have computed 
the ratio $\xi_2^{\rm LC}/\xi_0^{\rm LC}$ with the following 
parameterization for the bias instead of (\ref{defbz}),
\begin{equation}
  b(z)=\alpha+\bigl(b(z_*)-\alpha\bigr)
  \biggl({D_1(z_*)\over D_1(z)}\biggr)^\beta.
\label{defbzz}
\end{equation}
This alternation does not change the result significantly.
These situation suggest that precise measurements of 
correlation functions might offer a unique and useful 
test of the cosmological model of our universe, however, 
further investigation should be needed for the definite conclusion.

\section{Summary and Conclusions}
In the present paper we have revisited the feasibility of the geometric 
distortion focusing on the ambiguous factor of the evolution of bias. 
We have derived the useful formulas for the correlation function 
and the power spectrum which incorporate the light-cone effect 
and the redshift-space distortions in the rigorous manner by defining 
the estimators for the anisotropic correlation function and the 
power spectrum. The final expressions are consistent with those 
presented in the previous paper by Suto, Magira, and Yamamoto (2000).
Our investigation makes clear how the theoretical two-point correlation 
function and the power spectrum in our paper correspond 
to estimators in data processing.

For a quasar sample like the 2dF survey, whose distribution is extended 
to high-redshift, the light-cone effect becomes very important. 
In the correlation functions, the light-cone effect is 
incorporated by averaging a local correlation function over the 
redshift with the weight factor $W(z)$. With this formula we have 
investigated the theoretical predictions assuming a simple model of 
quasar sample which roughly match with the 2dF survey. 
Adopting a model for the evolution of 
bias (\ref{defbz}), which is phenomenologically parameterized in the 
simple form, we have examined the theoretical predictions for the 
two-point correlation functions. 
As pointed out by Ballinger et~al.(1996), the linear distortion 
is the dominant effect even at the high-redshift on the anisotropic 
correlation function $\xi_2^{\rm LC}$. The linear distortion is 
sensitive to the amplitude of bias, which limits the feasibility 
of the cosmological test with the geometric distortion. 

We have shown that the predicted correlation functions on the
light-cone are sensitive to the amplitude of bias $b(z_*)$ at 
the mean redshift $z_*$, and rather insensitive to the other 
parameters which specify the speed of its redshift-evolution
due to the averaging process from the light-cone effect. 
The characteristic correlation length of the isotropic 
correlation function can determine the mean amplitude of the 
bias, though we need to assume the value of $\sigma_8$.
If we take the discussions in subsection 3.3 seriously, the quasar 
correlation function is consistent with the $\Lambda$CDM model 
with the amplitude of the bias being $1.0\sim 2.0$ at $z_*=1.3$. 
We have also shown that $\xi^{\rm LC}_2(R)/\xi^{\rm LC}_0(R)$ is 
only sensitive
to the mean amplitude of bias, not to the speed of its
redshift-evolution. Taking the constraint on the mean amplitude 
of the bias from the characteristic correlation length into account,
the measurement of $\xi_2(R)/\xi_0(R)$ can
be a useful tool for 
testing the cosmological model of the universe. Conversely if 
the cosmological model is fixed, the precise measurement of 
$\xi_2(R)$ and $\xi_0(R)$ will provide a 
clue for the bias model. 

Finally we mention about other possible effects which affect 
measurements of statistical quantities. In general, the sampling 
noise and the shot noise limit the precise measurements of the 
statistical quantities.  The shape of a survey area might 
affect a proper estimation of the correlation functions too. 
Those effects depend on observational strategies. We cannot 
conclude whether the 2dF quasar sample will detect the cosmological 
constant at present stage and a numerical approach is needed 
to examine those observational effects to draw definite conclusions. 
Furthermore, though we have assumed the time-independent cosmological 
constant and the deterministic bias, the possibilities of the 
time-dependent cosmological constant like the quintessential CDM 
model and the stochasticity of the QSO clustering bias would 
make the feasibility of the geometric test more complex 
(Yamamoto \& Nishioka 2000). 
N-body numerical simulations of a cosmological horizon size, 
which is the frontier of numerical cosmology (e.g., Pearce et al. 
2000; Hamana et al. 2000), is another useful approach to 
investigate the geometric test.

%

\section*{Acknowledgments}

We are grateful to Prof. Kojima for useful comments.
We thank Prof. Suto for useful discussion and suggestions. 
We also thank referee for useful comments, which helped improve
the present manuscript. A.T. acknowledges the JSPS Research 
Fellowship. Numerical computations were carried out in part on 
INSAM of Hiroshima University. This work is supported by the 
Inamori foundation and in part by the Grants-in-Aid program 
(11640280) by the Ministry of Education, Science, Sports and 
Culture of Japan.

\newpage

\vspace{5mm}

\appendix
\section{}
\def\bfk{{\bf k}}
\def\bfq{{\bf q}}
\newcommand{\bfkpara}{{ k}_{\scriptscriptstyle \|}}
\newcommand{\bfkperp}{{\bf k}_{\scriptscriptstyle \bot}}
\newcommand{\bfqpara}{{ q}_{\scriptscriptstyle \|}}
\newcommand{\bfqperp}{{\bf q}_{\scriptscriptstyle \bot}}
\newcommand{\bfrperp}{{\bf r}_{\scriptscriptstyle \bot}}
\def\bfx{{\bf x}}
\def\bfr{{\bf r}}
\def\bfsh{{{\bf s}_h}}
\def\hbfsh{{\hat{{\bf s}}_h}}
\def\sh{{s_h}}
\def\f{{f_{K}}}

In this Appendix we give an explicit derivation of the formula 
(\ref{Pcllk}).  Though the formula is presented in the paper 
(Suto et~al. 2000), the explicit derivation is not presented there. 
We believe that it is useful to present the way to derive the 
formula from the estimators $P_l(k)$ and $\xi_l(R)$ defined explicitly.
From the definition of the estimators, we write 
\begin{eqnarray}
 &&\xi^{\rm LC}_l(R)=\int
 {d\Omega_{\hat\bfR}\over 4\pi}\int d\bfs_1 \int d\bfs_2
 \bigl<F(\bfs_1)F(\bfs_2)\bigr>
   \delta^{(3)}(\bfs_1-\bfs_2-\bfR)\P_l({\hat\bfs}_h\cdot\hat\bfR)(2l+1).
\label{defxila}
\end{eqnarray}
With the aid of equation (\ref{synth}), $\bigl<F(\bfs_1)F(\bfs_2)\bigr>$ 
in (\ref{defxila}) reduces to
\begin{eqnarray}
  &&\bigl<F(\bfs_1)F(\bfs_2)\bigr>={\tilde n(s_1)\tilde n(s_2) 
  \bigl<\Delta(\bfs_1)\Delta(\bfs_2)\bigr> \over
  \int d\bfs \tilde n(s)^2}.
\end{eqnarray}
Instead of the variables $\bfs_1$ and $\bfs_2$, we introduce
\begin{eqnarray}
  &&\bfx=\bfs_1-\bfs_2,
\\
  &&\bfsh={1\over 2}(\bfs_1+\bfs_2),
\end{eqnarray}
then we have
\begin{eqnarray}
 &&\xi^{\rm LC}_l(R)=\bigl[4\pi \int ds s^2 \tilde n(s)^2\bigr]^{-1}
  \int{d\Omega_{\hat\bfR}\over 4\pi}\int d\bfx \int d\bfsh
  \tilde n(s_1)\tilde n(s_2)  \bigl<\Delta(\bfs_1)\Delta(\bfs_2)\bigr>
\nonumber
\\
 &&{\hspace{0.5cm}}\times
   \delta^{(3)}(\bfx-\bfR)\P_l({\hat\bfs}_h\cdot\hat\bfR)(2l+1),
\label{defxilb}
\end{eqnarray}
where $s_h=|\bfs_h|$ and $\hat s_h=\bfs_h/|\bfs_h|$.
Exact calculations for $\xi_0^{\rm LC}(R)$ and $P_0^{\rm LC}(k)$ 
are performed within linear theory of density perturbations
in references (Suto \& Yamamoto 1999; Nishioka \& Yamamoto 1999; 2000).
Though all the redshift distortions are not incorporated simultaneously
in those works, however, those investigations have shown the validity 
for using the distant observer approximation. Then we here use the 
distant-observer approximation:
\begin{eqnarray}
  &&\tilde n(s_1)\tilde n(s_2)\simeq \tilde n(\sh)^2,
\\
  &&\bigl<\Delta(\bfs_1)\Delta(\bfs_2)\bigr>\simeq 
  \xi(\sh,x,{\hbfsh}\cdot\hat\bfx),
\end{eqnarray}
where $\hat\bfx=\bfx/|\bfx|$ and $x=|\bfx|$.
Here we have approximated that $\bigl<\Delta(\bfs_1)\Delta(\bfs_2)\bigr>$
is a function of $\sh$, $x$, and $\hbfsh\cdot\hat\bfx$.

Now we consider the expression for  $\xi(\sh,x,\hbfsh\cdot\hat\bfx)$.
We introduce the following coordinates to describe real universe,
\begin{eqnarray}
  &&ds^2=a(\eta)^2\Bigl(-d\eta^2+dr^2+\f(r)^2d\Omega_{(2)}^2\Bigr),
\end{eqnarray}
where 
\begin{eqnarray}
\label{eq: comoving D_A}
 \f(r)=\,\left\{
\begin{array}{lr}
\sin{(\sqrt{K}\,r)}/\sqrt{K} & (K>0) \\
r & (K=0) \\
\sinh{(\sqrt{-K}\,r)}/\sqrt{-K} & (K<0) \\
\end{array}
\right. , 
\end{eqnarray}
with the spatial curvature $K=H_0^2(\Omega_0+\Omega_\Lambda-1)
=-H_0^2\Omega_K$. Then the number conservation means
\begin{eqnarray}
  &&n(s,\gamma) s^2ds=n^{\rm R}(r,\gamma) \f(r)^2 dr,
\label{ncons}
\end{eqnarray}
where $n^{\rm R}(r,\gamma)$ denotes the number density field
in real space. Introducing the mean number density 
$\tilde n(r)^{\rm R}$ at the distance $r$ in real space, 
we write
\begin{equation}
  n^{\rm R}(r,\gamma)=\tilde n^{\rm R}(r)(1+\Delta^{\rm R}(r,\gamma)),
\label{deltanR}
\end{equation}
where $\Delta^{\rm R}(r,\gamma)$ denotes the density contrast in real 
space. Combining (\ref{ncons}) and (\ref{deltanR}), we have
\begin{eqnarray}
  &&\tilde n(s)s^2ds=\tilde n^{\rm R}(r)\f(r)^2 dr,
\\
  &&\tilde n(s)\Delta(s,\gamma)s^2ds
  =\tilde n^{\rm R}(r)\Delta^{\rm R}(r,\gamma)\f(r)^2 dr,
\end{eqnarray}
and
\begin{eqnarray}
  &&\bigl<\Delta(\bfs_1)\Delta(\bfs_2)\bigr>
  =\bigl<\Delta^{\rm R}(\bfr_1)\Delta^{\rm R}(\bfr_2)\bigr>,
\label{DD}
\end{eqnarray}
where $\bfr$ denotes the point specified by $r$ and $\gamma$.
When the distance $|\bfr_1-\bfr_2|$ is small compared with the size of the
survey volume, we may write 
\begin{eqnarray}
  &&\bigl<\Delta^{\rm R}(\bfr_1)\Delta^{\rm R}(\bfr_2)\bigr>
\nonumber
\\
  &&\hspace{0cm}
  =\int {d\bfq\over (2\pi)^3}e^{i\bfq\cdot(\bfr_1-\bfr_2)} 
  P(\bfqpara,\bfqperp,z)
\nonumber
\\
  &&\hspace{0cm}
  =\int {d\bfqpara\bfqperp  \over (2\pi)^3}
  \exp\bigl[{i\bfqpara(r_1-r_2)
            +i\bfqperp\cdot(\bfr_{\perp,1}-\bfr_{\perp,2})}\bigr] 
  P(\bfqpara,\bfqperp,z),
\label{DDD}
\end{eqnarray}
where $P(\bfqpara,\bfqperp,z)$ denotes the power spectrum 
of cosmological objects in real space (including the effect of the 
peculiar motion) on a constant time hypersurface at redshift $z$,
and we introduced $\bfr=(r,\bfrperp),~\bfq=(\bfqpara,\bfqperp)$,
where the subscript $||$ denotes the component of the line of sight 
and $\perp$ denotes the components perpendicular to the vector of 
the line of sight. 
Here the redshift $z$ in $P(\bfqpara,\bfqperp,z)$ should be 
understood as a function of the comoving distance 
$r_h=(r_1+r_2)/2$. 
Even in the case of the open universe, we can expand the
correlation function in terms of the usual Fourier modes,
as in (\ref{DDD}), because we are considering the case the 
distance $|\bfr_1-\bfr_2|$ is sufficiently small compared 
with the horizon (curvature) scale. 
Under the distant observer approximation we have
\begin{eqnarray}
  &&\bfr_{\perp,1}-\bfr_{\perp,2}
  ={\f(r_h)\over s_h}(\bfs_{\perp,1}-\bfs_{\perp,2}),
\\
  &&r_1-r_2={dr_h\over ds_h}(s_1-s_2),
\end{eqnarray}
where $s_h=|\bfs_1+\bfs_2|/2$.
Introducing the new variables,
\begin{eqnarray}
  &&\bfkperp={\f(r_h)\over s_h}\bfqperp=\cperp\bfqperp,
\\
  &&\bfkpara={dr_h\over ds_h}\bfqpara=\cpara\bfqpara,
\end{eqnarray}
we have
\begin{eqnarray}
  &&\bigl<\Delta(\bfs_1)\Delta(\bfs_2)\bigr>
  ={1\over \cperp^2 \cpara}\int {d\bfk\over (2\pi)^3}
  e^{i\bfk\cdot(\bfs_1-\bfs_2)} 
P\bigl(\bfqpara\rightarrow{\bfkpara\over \cpara},
           \bfqperp\rightarrow{\bfkperp\over \cperp},z\bigr),
\end{eqnarray}
where we used (\ref{DDD}) and (\ref{DD}). 
Here the redshift $z$ should be understood
as a function of $s_h$ instead of $r_h$. Inserting this equation into 
(\ref{defxilb}), we have 
\begin{eqnarray}
   &&\xi^{\rm LC}_l(R)=\bigl[4\pi \int dss^2 \tilde n(s)^2\bigr]^{-1}
  i^l\int {d \Omega_{\hat s_h}}
  \int ds_h s_h^2 \tilde n(s_h)^2 {1\over \cperp^2\cpara}
  \int {d\bfk\over (2\pi)^3} j_l(kR) 
\nonumber
\\
  &&\hspace{0.5cm} \times 
  P\bigl(\bfqpara\rightarrow{\bfkpara\over \cpara},
           \bfqperp\rightarrow{\bfkperp\over \cperp},z\bigr)
  \P_l({\hat\bfs}_h\cdot\hat\bfk)(2l+1),
\end{eqnarray}
where we used the mathematical formulas,
\begin{eqnarray}
  &&e^{i\bfk\cdot\bfx}= 4\pi \sum_{l} \sum_{m=-l}^{l}
  i^l j_{l}(kx) Y^{m*}_{l}(\Omega_{\hat\bfk})
  Y^m_{l}(\Omega_{\hat\bfx}),
\label{mathA}
\\
  &&\P_l({\hat\bfs}_h\cdot\hat\bfk)={4\pi\over 2l+1} \sum_{m=-l}^{l}
   Y^m_{l}(\Omega_{{\hat\bfs}_h})Y^{m*}_{l}(\Omega_{\hat\bfk}).
\label{mathB}
\end{eqnarray}
By introducing $\mu=\bfkpara/k$, we have
\begin{eqnarray}
   &&\xi^{\rm LC}_l(R)=\bigl[\int dss^2 \tilde n(s)^2\bigr]^{-1}
  i^l\int ds_h s_h^2 \tilde n(s_h)^2 
  {2l+1\over \cperp^2\cpara}
  \int_0^\infty {dk k^2\over 2\pi^2} j_l(kR)\int_0^1 d\mu  
\nonumber
\\
  &&\hspace{0.2cm}
  \times P\bigl(\bfqpara\rightarrow{k\mu\over \cpara},
           |\bfqperp|\rightarrow{k\sqrt{1-\mu^2}\over \cperp},z\bigr)
  \P_l(\mu).
\end{eqnarray}
The transformation (\ref{invtr}) leads
\begin{eqnarray}
  &&P^{\rm LC}_l(k)=\bigl[\int dss^2 \tilde n(s)^2\bigr]^{-1}
    \int ds_h s_h^2 \tilde n(s_h)^2 
  {2l+1\over \cperp^2\cpara}
  \int_0^1 d\mu   \P_l(\mu)
\nonumber
\\
  &&\hspace{0.2cm}
  \times P\bigl(\bfqpara\rightarrow{k\mu\over \cpara},
           |\bfqperp|\rightarrow{k\sqrt{1-\mu^2}\over \cperp},z\bigr).
\end{eqnarray}

The effects of the linear distortion, the 
nonlinear effect, and the finger of god effect have been investigated 
for the two-point statistics on a constant time hypersurface 
(Peacock \& Dodds 1994; 1996, Ballinger et~al. 1996, 
Magira, Jing, \& Suto 2000).
We adopt the following formula, 
\begin{eqnarray}
  &&P(\bfqpara,|\bfqperp|,z)
  =\Bigl\{1+\beta(z)\Bigl({\bfqpara\over q}\Bigr)^2\Bigr\}^2
  b(q,z)^2P_{\rm mass}(q,z)D\Bigl[\bfqpara \sigma_P(z)\Bigr],
\label{appendixP}
\end{eqnarray}
with $q=\sqrt{\bfqpara^2+|\bfqperp|^2}$, mass power spectrum 
$P_{\rm mass}(q,z)$, the bias factor $b(q,z)$, the
damping factor $D[\bfqpara \sigma_P(z)]$, and the $\beta(z)$
factor defined by equation (\ref{betafactor}).
Here we have assumed that the number density fluctuation of 
cosmological objects is simply proportional to the mass density 
fluctuation in real space by introducing the linear bias factor $b(q,z)$.  
Then we obtain the expression (\ref{Pcllk}), using the relation 
$dN=ds s^2 \tilde n(s)$.

For convenience we summarize the explicit formulas for the factors 
$\cpara$ and $\cperp$. Throughout this paper we assume that the 
correct model of our universe is the CDM model. 
Then the comoving distance in the real space is expressed as
\begin{equation}
  r(z)={1\over H_0} \int_0^z 
  {dz' \over \sqrt{\Omega_0(1+z')^3+\Omega_K(1+z')^2+\Omega_\Lambda}}.
\nonumber
\\
\end{equation}
With this formula, $\cperp$ is given by
\begin{equation}
  \cperp={\f(r(z))\over s(z)},
\label{defcperpp}
\end{equation}
and 
\begin{eqnarray}
 \cpara= \displaystyle{{1\over \sqrt{\Omega_0+\Omega_K(1+z)^{-1}+
\Omega_\Lambda(1+z)^{-3}}} },
\end{eqnarray}
for the case $s=s_{Ed}(z)$, and 
\begin{eqnarray}
\cpara= \displaystyle{{1\over \sqrt{\Omega_0(1+z)^{3}+\Omega_K(1+z)^{2}+
  \Omega_\Lambda}} },
\end{eqnarray}
for the case $s=s_{Hb}(z)$, respectively.

\bigskip
\bigskip

\label{lastpage}
\bsp
\appendix

\newpage
\vspace{2cm}
\begin{center}
{\large \bf FIGURE CAPTIONS}
\end{center}
\noindent
{Fig.~1---
A sketch to illustrate variables defined in this paper.
}

\vspace{0.5cm}
\noindent{Fig.~2---
 The weight function $W(z)$ for the quasar sample in section 3.1. 
 The dashed line shows the case adopting $s=s_{Hb}(z)$, and the 
 solid line does the case $s=s_{Ed}(z)$.
}

\vspace{0.5cm}
\noindent{Fig.~3---
 Power spectra on a constant-time hypersurface.
 The upper panels show $P_0^{\rm crd}(k,z)$ in unit of $h^{-3}{\rm Mpc}^3$, 
 the middle panels show $P_2^{\rm crd}(k,z)$, and the lower panels show 
 the ratio $P_2^{\rm crd}(k,z)/P_0^{\rm crd}(k,z)$.
 The left panels show the case $z=0.5$, the center $z=2$, and
 the right $z=4$. We assumed the $\Lambda$CDM model with  
 $\Omega_0=0.3$, $\Omega_\Lambda=0.7$, $h=0.7$, and $\sigma_8=1.0$.
 In this figure the bias parameter is fixed as $\alpha=0.5$, 
 $\beta=1$, and $b(z_*=2)=2$. Here $s=s_{Hb}(z)$ is adopted.
 In each panel, the solid line plots the case obtained from the 
 formula (15), while the dotted line plots the case 
 when the nonlinear and the finger of God effects are neglected, 
 and the dashed line plots the case when the distortion due to the 
 peculiar motion of the sources and the nonlinear effect are 
 neglected. The dashed line shows the case that only the geometric 
 distortion is taken into account.
}

\vspace{0.5cm}
\noindent{Fig.~4---
Same as Fig.3 but with adopting the cosmological
redshift-space $s=s_{Ed}(z)$.
}

\vspace{0.5cm}
\noindent{Fig.~5---
Contour of the characteristic correlation length $R_c$
 on the bias-parameter plane $\beta$ and $b(z_*)$ with $z_*=1.3$. 
 We fixed $\alpha=0$ in the left panels, $\alpha=0.5$ in the center
 panels, and $\alpha=1.0$ in the right panels. 
 The upper panels show the case when the $\Lambda$CDM model, 
 $\Omega_0=0.3$, $\Omega_\Lambda=0.7$, $h=0.7$, $\sigma_8=1.0$,
 is assumed as the cosmological model of the real space. While 
 the lower panels show the case that the open CDM model with
 $\Omega_0=0.3$, $\Omega_\Lambda=0.0$, $h=0.7$, $\sigma_8=1.0$,
 is adopted as the real space. We adopted the cosmological 
 redshift space $s=s_{Ed}(z)$. In each panel, the lines show the 
 contour of the levels, $R_c=10$, $8$, $6$, $4$, $3~h^{-1}{\rm Mpc}$,
 from right to left. The solid lines are the contours 
 $R_c=6$ and $4~h^{-1}{\rm Mpc}$.
}

\vspace{0.5cm}
\noindent{Fig.~6---
Contour of $\xi_2^{\rm LC}(R)/\xi_0^{\rm LC}(R)$ in the
 $\Lambda$CDM model with fixed $R=5h^{-1}$Mpc (upper panels), 
 $R=10h^{-1}$Mpc (middle panels), and $R=20h^{-1}$Mpc (lower panels). 
 Similar to figure 5, we chose $\alpha=0$ (left panels), $0.5$ (center 
 panels),  and $1.0$ (right panels). In this figure all the distortion
 effects are taken into account. The cosmological parameters, 
 $\Omega_0=0.3$, $\Omega_\Lambda=0.7$, $h=0.7$, $\sigma_8=1.0$,
 are taken, and we adopted the cosmological redshift space $s=s_{Ed}(z)$.
}

\vspace{0.5cm}
\noindent{Fig.~7---
Contour of $\xi_2^{\rm LC}(R)/\xi_0^{\rm LC}(R)$ with fixed $R=10h^{-1}$Mpc
on the bias-parameter plane $\beta$ and $b(z_*)$. We fixed $\alpha=0$
(left panels), $0.5$ (center panels), and $1.0$ (right panels). 
The upper panels show the case that the peculiar motion 
of sources is neglected, that is, the linear 
distortion and the nonlinear and the finger of god effects are omitted. 
The upper panels show the effect only from the geometric distortion.
The middle panels show the case the geometric distortion and 
the linear distortion are incorporated but the nonlinear and 
the finger of god effects are omitted. The lower panels show 
the case all the distortion effects are incorporated. 
In this figure the $\Lambda$CDM model with the same cosmological 
parameters as those in figure 6 is assumed.
}

\vspace{0.5cm}
\noindent{Fig.~8---
Contour of $\xi_2^{\rm LC}(R)/\xi_0^{\rm LC}(R)$ 
with fixed $R=10h^{-1}$Mpc for various cosmological models.
(a) $\Lambda$CDM model with $\Omega_0=0.3$, $\Omega_\Lambda=0.7$;
(b) $\Lambda$CDM model with $\Omega_0=0.4$, $\Omega_\Lambda=0.6$;
(c) open CDM model with $\Omega_0=0.3$, $\Omega_\Lambda=0.0$;
(d) open CDM model with $\Omega_0=0.4$, $\Omega_\Lambda=0.0$.
In each model we adopted $h=0.7$ and $\alpha=0.5$. 
The contour lines are $-0.8$, $-0.5$, $-0.3$ from left to right for the 
$\Lambda$CDM models. While the contour levels are
$-0.5$, $-0.3$, $-0.1$ for the open CDM models from left to right.
The region between the dashed lines in each panel satisfies $4\leq R_c \leq6$.
Here $\sigma_8$ is determined from the cluster abundance 
(Kitayama \& Suto 1997).
We assumed the cosmological redshift space with $s=s_{Ed}(z)$.
}

\vspace{0.5cm}
\noindent{Fig.~9---
Same figure as figure 8 but with fixed $R=20h^{-1}$Mpc.
The contour-levels drawn by solid lines are shown in each panel.
}
\end{document}

\newpage
 \begin{figure}
 \begin{center}
 \leavevmode
 \epsfile{file=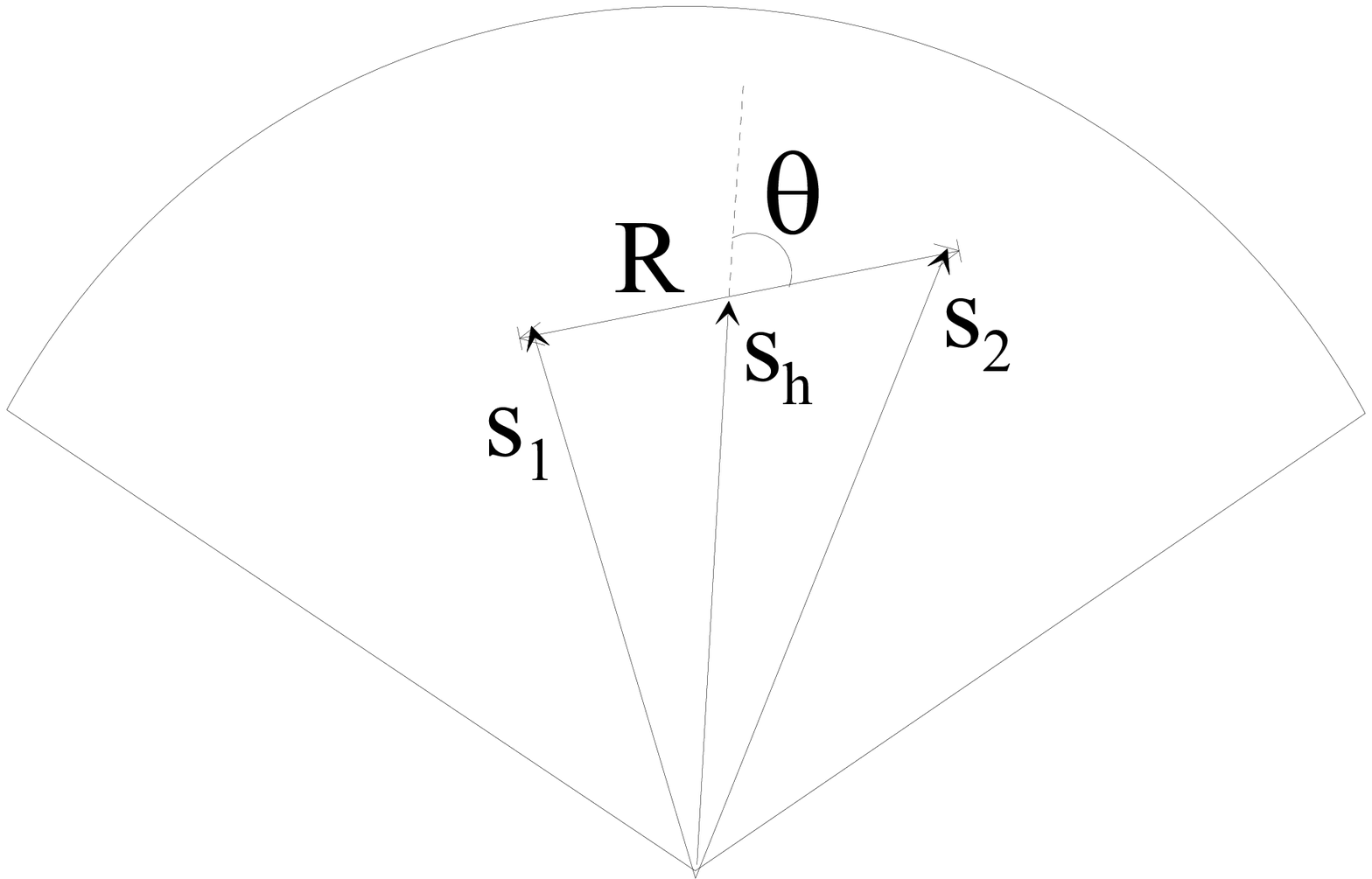,width=15cm}
 \end{center}
 \caption{
}
 \end{figure}

\newpage
 \begin{figure}
 \begin{center}
 \leavevmode
 \epsfile{file=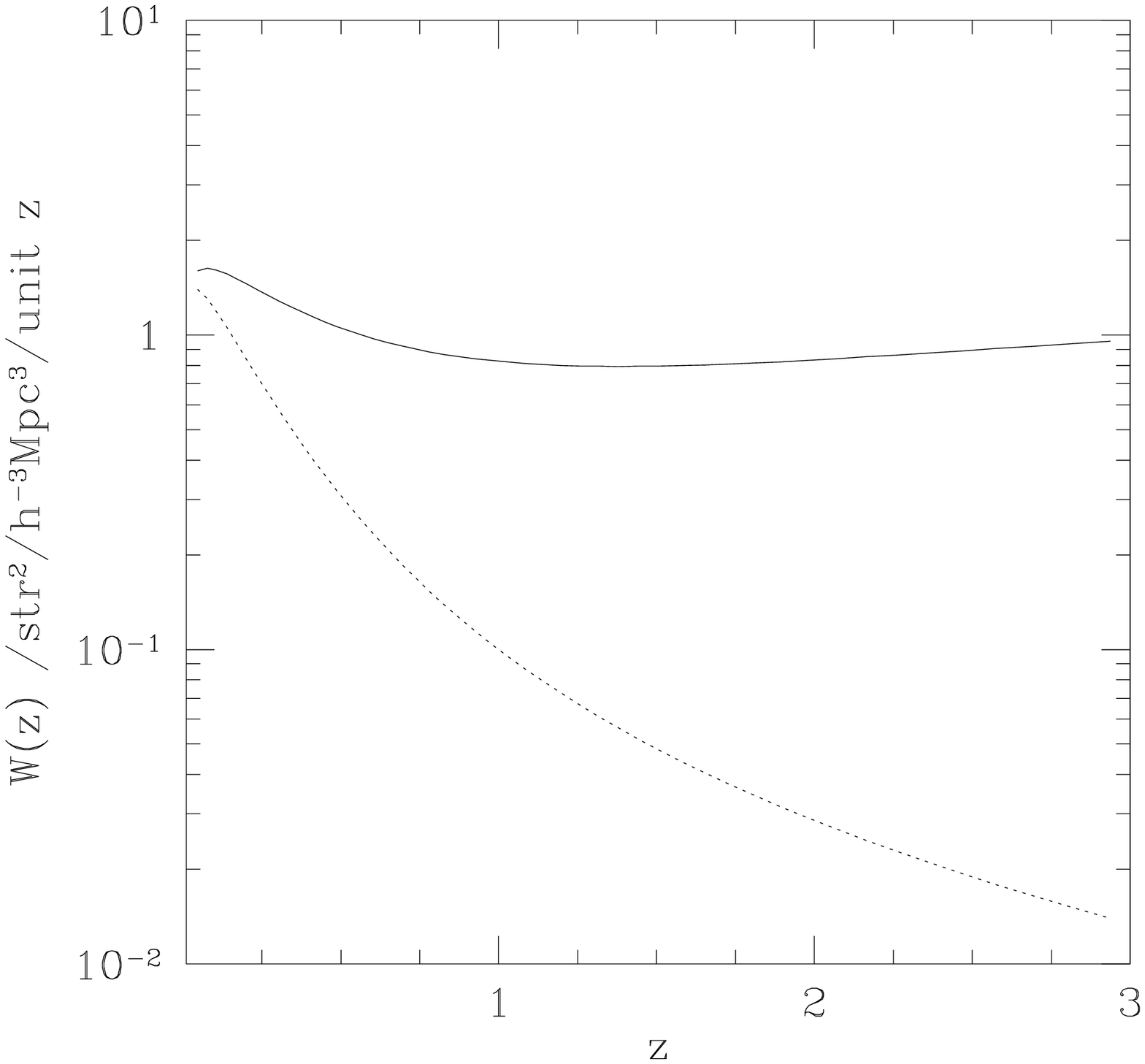,width=15cm}
 \end{center}
 \caption{
}
 \end{figure}

\newpage
 \begin{figure}
 \begin{center}
 \leavevmode
 \epsfile{file=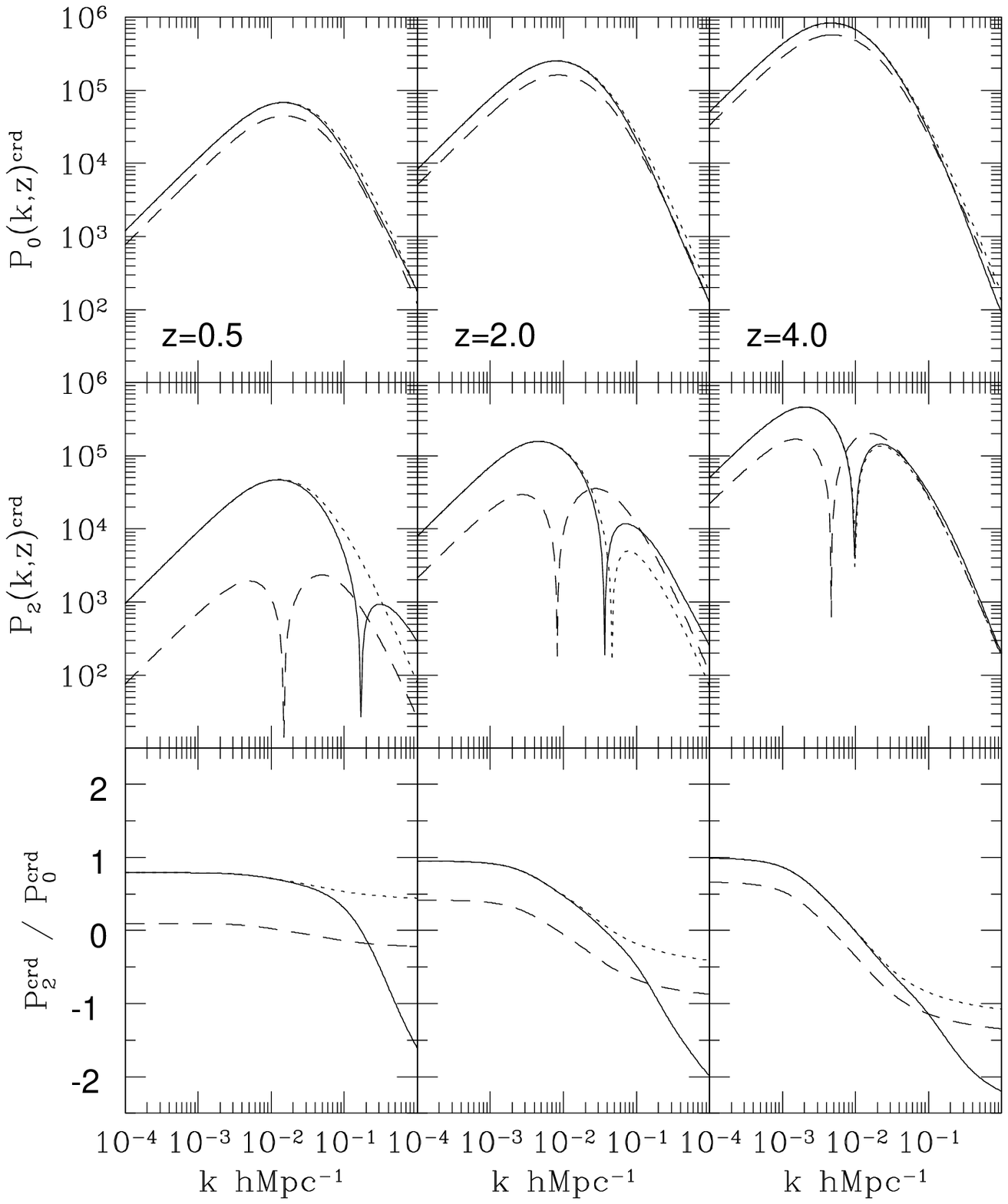,width=18cm}
 \end{center}
 \caption{
}
 \end{figure}

\newpage
 \begin{figure}
 \begin{center}
 \leavevmode
 \epsfile{file=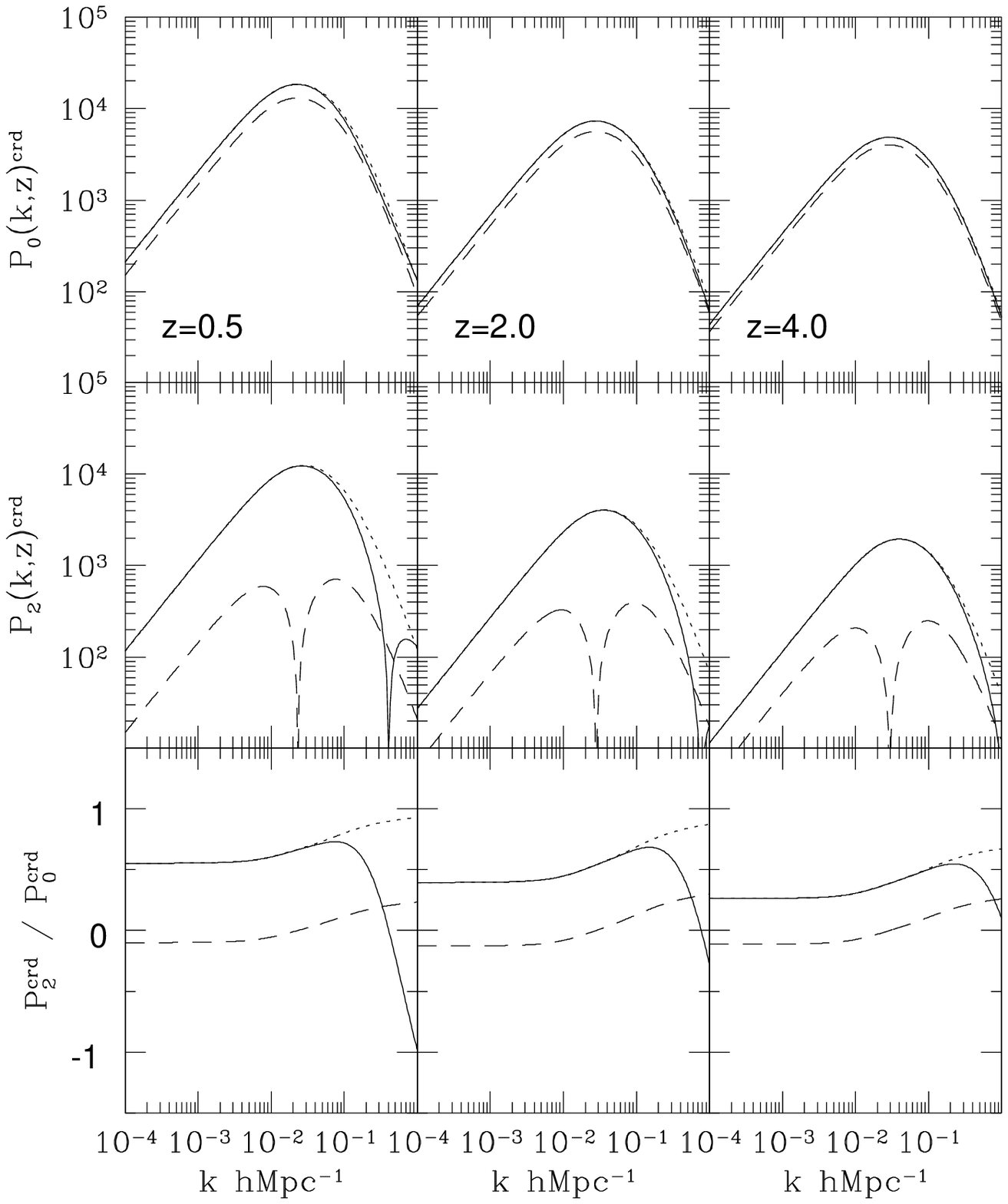,width=18cm}
 \end{center}
 \caption{
}
 \end{figure}

\newpage
 \begin{figure}
 \begin{center}
 \leavevmode
 \epsfile{file=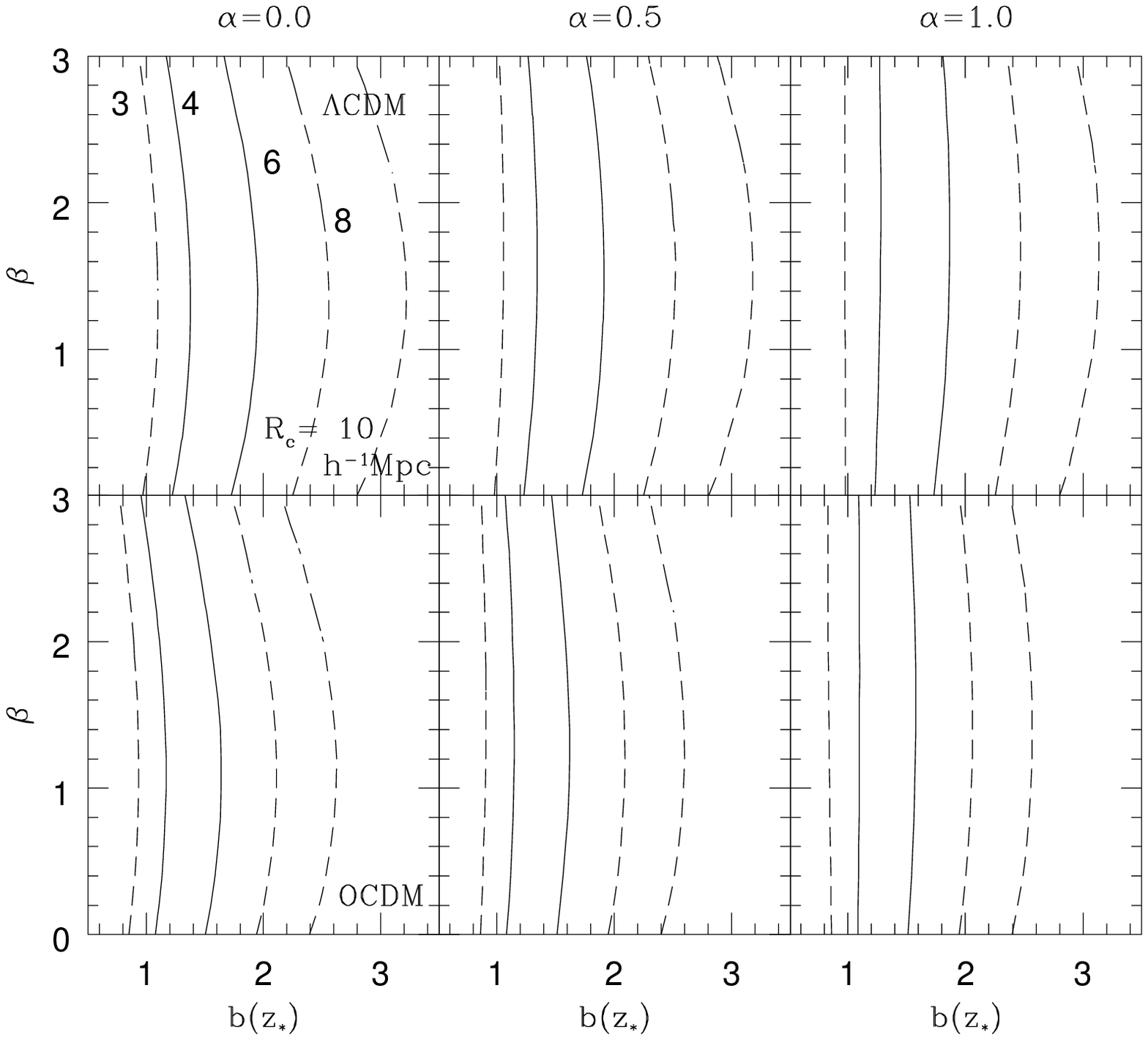,width=17cm}
 \end{center}
 \caption{
}
\end{figure}

\newpage
 \begin{figure}
 \begin{center}
 \leavevmode
 \epsfile{file=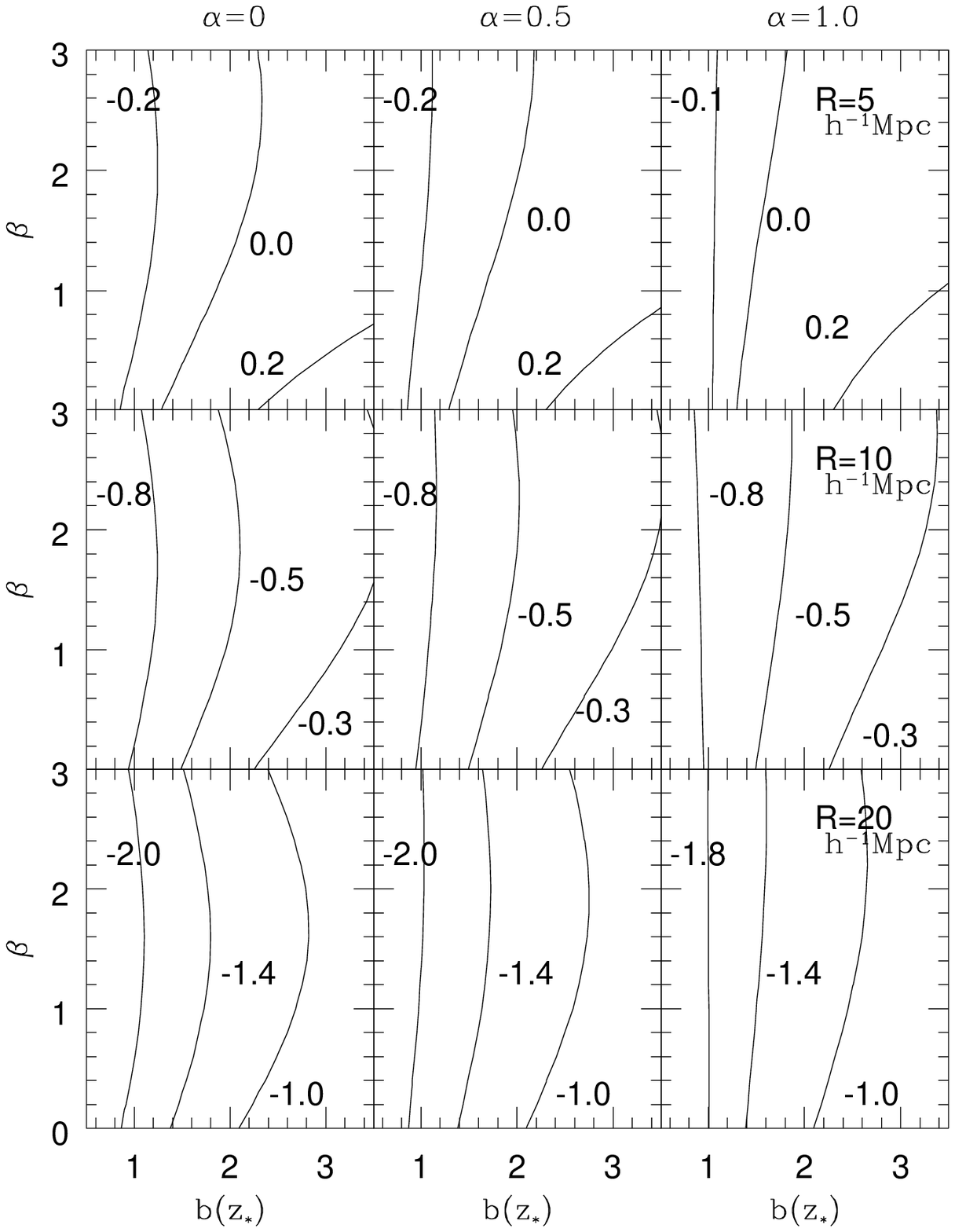,width=18cm}
 \end{center}
 \caption{
}
 \end{figure}

\newpage
 \begin{figure}
 \begin{center}
 \leavevmode
 \epsfile{file=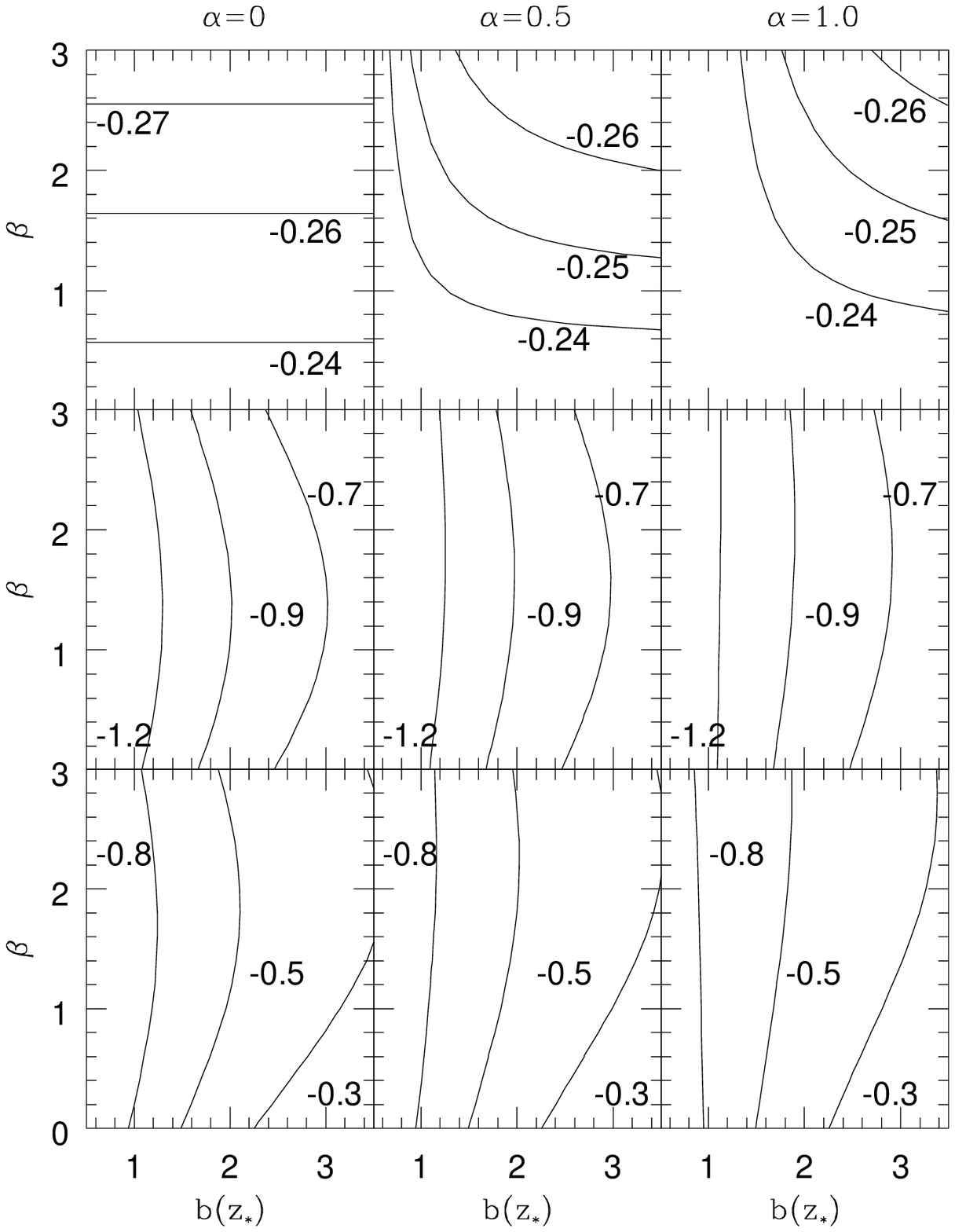,width=18cm}
 \end{center}
 \caption{
}
 \end{figure}

\newpage
 \begin{figure}
 \begin{center}
 \leavevmode
 \epsfile{file=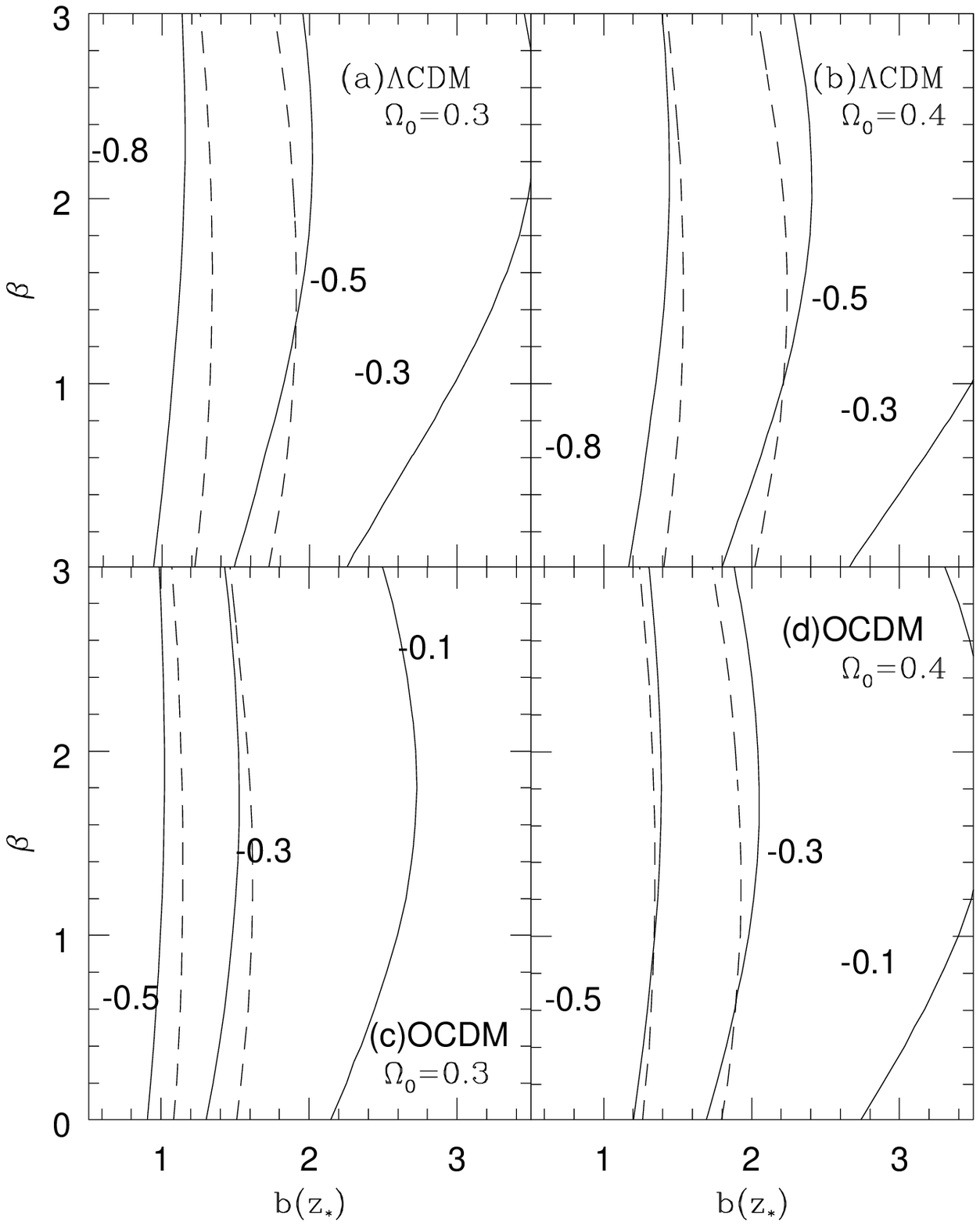,width=18cm}
 \end{center}
 \caption{
}
 \end{figure}

\newpage
 \begin{figure}
 \begin{center}
 \leavevmode
 \epsfile{file=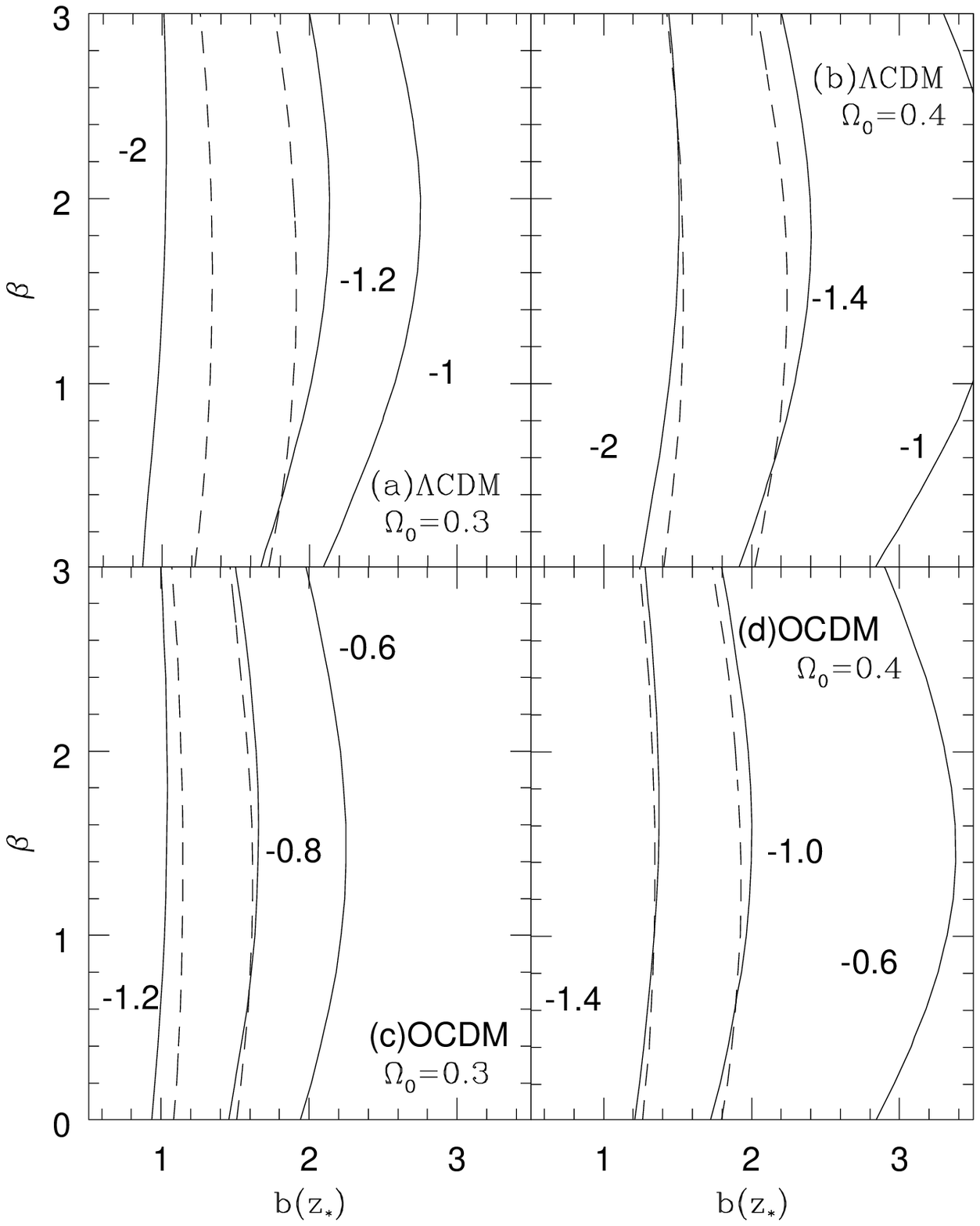,width=18cm}
 \end{center}
 \caption{
}
 \end{figure}
\end{document}

%% file: mnmacro.tex
\ifoldfss
  \ifCUPmtlplainloaded \else
    \NewTextAlphabet{textbfit} {cmbxti10} {}
    \NewTextAlphabet{textbfss} {cmssbx10} {}
    \NewMathAlphabet{mathbfit} {cmbxti10} {} 
    \NewMathAlphabet{mathbfss} {cmssbx10} {} 
  \fi
  \ifAMStwofonts
    \ifCUPmtlplainloaded \else
      \NewSymbolFont{upmath} {eurm10}
      \NewSymbolFont{AMSa} {msam10}
      \NewMathSymbol{\upi}     {0}{upmath}{19}
      \NewMathSymbol{\umu}     {0}{upmath}{16}
      \NewMathSymbol{\upartial}{0}{upmath}{40}
      \NewMathSymbol{\leqslant}{3}{AMSa}{36}
      \NewMathSymbol{\geqslant}{3}{AMSa}{3E}

      \let\leq=\leqslant 
       
    \fi
  \fi
\fi 

\ifnfssone
  \newmathalphabet{\mathit}
  \addtoversion{normal}{\mathit}{cmr}{m}{it}
  \addtoversion{bold}{\mathit}{cmr}{bx}{it}
  \newmathalphabet{\mathbfit} 
  \addtoversion{normal}{\mathbfit}{cmr}{bx}{it}
  \addtoversion{bold}{\mathbfit}{cmr}{bx}{it}
  \newmathalphabet{\mathbfss} 
  \addtoversion{normal}{\mathbfss}{cmss}{bx}{n}
  \addtoversion{bold}{\mathbfss}{cmss}{bx}{n}
  \ifAMStwofonts
    \ifCUPmtlplainloaded \else
      \UseAMStwoboldmath
      \makeatletter
      \new@mathgroup\upmath@group
      \define@mathgroup\mv@normal\upmath@group{eur}{m}{n}
      \define@mathgroup\mv@bold\upmath@group{eur}{b}{n}
      \edef\UPM{\hexnumber\upmath@group}
      \new@mathgroup\amsa@group
      \define@mathgroup\mv@normal\amsa@group{msa}{m}{n}
      \define@mathgroup\mv@bold\amsa@group{msa}{m}{n}
      \edef\AMSa{\hexnumber\amsa@group}
      \makeatother
      \mathchardef\upi="0\UPM19
      \mathchardef\umu="0\UPM16
      \mathchardef\upartial="0\UPM40
      \mathchardef\leqslant="3\AMSa36
      \mathchardef\geqslant="3\AMSa3E

      \let\leq=\leqslant 

    \fi
  \fi
\fi 

\ifnfsstwo
  \DeclareMathAlphabet{\mathbfit}{OT1}{cmr}{bx}{it}
  \SetMathAlphabet\mathbfit{bold}{OT1}{cmr}{bx}{it}
  \DeclareMathAlphabet{\mathbfss}{OT1}{cmss}{bx}{n}
  \SetMathAlphabet\mathbfss{bold}{OT1}{cmss}{bx}{n}
  \ifAMStwofonts
    \ifCUPmtlplainloaded \else
      \DeclareSymbolFont{UPM}{U}{eur}{m}{n}
      \SetSymbolFont{UPM}{bold}{U}{eur}{b}{n}
      \DeclareSymbolFont{AMSa}{U}{msa}{m}{n}
      \DeclareMathSymbol{\upi}{0}{UPM}{"19}
      \DeclareMathSymbol{\umu}{0}{UPM}{"16}
      \DeclareMathSymbol{\upartial}{0}{UPM}{"40}
      \DeclareMathSymbol{\leqslant}{3}{AMSa}{"36}
      \DeclareMathSymbol{\geqslant}{3}{AMSa}{"3E}

      \let\leq=\leqslant 

    \fi
  \fi
\fi 

\ifCUPmtlplainloaded \else
  \ifAMStwofonts \else 
    \def\upi{\pi}
    \def\umu{\mu}
    \def\upartial{\partial}
  \fi
\fi